\documentclass[showpacs,preprintnumbers,amsmath,amssymb,superscriptadress,prb,letterpaper,twocolumn,aps,epsf]{revtex4}
\usepackage{times}
\usepackage{dcolumn}
\usepackage{amsmath}
\usepackage{graphicx}%

\def\comment#1{}

\newcommand{\be}{\begin{equation}}
\newcommand{\ee}{\end{equation}}
\newcommand{\ba}{\begin{eqnarray}}
\newcommand{\ea}{\end{eqnarry}}
\newcommand{\bastar}{\begin{eqnarray*}}
\newcommand{\eastar}{\end{eqnarray*}}

\newcommand{\beg}{\begin{eqnarray}}
\newcommand{\eee}{\end{eqnarray}}

\newcommand{\avg}[1]{\left<#1\right>}

\def\cm#1{}

 \newcommand{\bos}[1]{\boldsymbol{#1}}
 
\newcommand{\ex}[1]{\mathrm{e}^{#1}}

\newcommand{\du}{\mathrm{d}}  

\newcommand{\newatop}[2]{\genfrac{}{}{0pt}{}{#1}{#2}}

\begin{document}
 
\title{Preemptive vortex-loop proliferation in multicomponent interacting
  Bose--Einstein condensates} \author{E. K. Dahl$^{1}$,
  E. Babaev$^{2,3}$, S. Kragset$^{1}$, and A. Sudb{\o}$^{1}$ }
\affiliation{
  $^{1}$ Department of Physics, Norwegian University of Science and Technology, N-7491 Trondheim, Norway\\
  $^{2}$ Physics Department, University of Massachusetts, Amherst MA 01003, USA\\
  ${}^3$ Department of Theoretical Physics, The Royal Institute of
  Technology 10691 Stockholm, Sweden}

\date{\today}

\begin{abstract}
We use analytical arguments and large-scale Monte Carlo calculations to investigate 
the nature of the phase transitions between distinct complex superfluid phases in a 
two-component Bose--Einstein condensate when a non-dissipative drag between the two 
components is being varied. We focus on understanding the role of topological defects  
in various phase transitions and develop vortex-matter arguments allowing an analytical 
description of the phase diagram. We find the behavior of fluctuation induced vortex 
matter to be much more complex and substantially different from that of single-component 
superfluids. We propose and investigate numerically a novel drag-induced ``preemptive vortex
loop proliferation'' transition. Such a transition may be a quite generic feature in many 
multicomponent systems where symmetry is restored by a gas of several kinds of competing 
vortex loops.
\end{abstract}


\maketitle
\section{\label{Intro}Introduction}
Natural generalizations of many superfluid phenomena are possible in
mixtures of independently conserved multicomponent Bose--Einstein
condensates with intercomponent current-current interactions. The
topic was first investigated in the context of ${}^4{\rm He}- {}^3{\rm
  He}$ mixtures \cite{AB,AB-2}, where it is possible to attain only a
limited range of parameters. The recent progress in atomic
Bose--Einstein condensates (BEC) has made it possible to access a much
wider range of regimes and explore novel superfluid phases which can
arise in such mixtures. For this reason, there has been much  interest in 
a generic example of an interacting BEC mixture, namely a $U(1)\times
U(1)$-symmetric system with current-current interactions. One of the
novel aspects of the superfluid physics in such a system is the
possibility of a phase transition at a sufficiently strong
current-current interaction to a state of paired superfluid where the
only broken $U(1)$ symmetry is associated with order only in the phase
sum \cite{amh1}. The other discussed example (which does not fall
within the framework of Galilean-invariance based argument \cite{AB})
is a phase transition for bosons on an optical lattice to a state
where one species of bosons pair with holes of the other species, and
thereby retaining order only in the phase difference
\cite{Kuklov1,Kuklov,amh1}.
\par
These transitions were investigated numerically in great detail in the
$J$-current model in Ref.  \onlinecite{amh2} using the worm-algorithm
\cite{worm-algorithm}. This numerical study, combined with mean-field 
arguments, revealed the interesting fact that
with increasing current-current interaction, the usual second-order
superfluid phase transition is altered to a first order phase
transition \cite{amh2}. In the free energy functional, the
current-current interaction is consistent with $U(1)\times U(1)$
symmetry and the transition should therefore be associated with a
proliferation of interacting vortex loops where all vortex-loop
segments of the system interact with each other through a Coulomb
potential. Existing theories of proliferation of such defects,
however, always lead to a second-order superfluid phase transition
\cite{blowup}. This indicates that in this system we are faced with a
novel scenario for thermally driven spontaneous vortex-loop
proliferation, the detailed investigation of which is the goal of the
present work.
\par
To describe the behavior of the system undergoing these phase
transitions as proliferation of vortex loops in a two-component
condensate, we propose a scenario of a ``preemptive vortex-loop
proliferation''.  This scenario in particular allows us to estimate the
characteristic critical couplings (or equivalently, critical
temperatures) and provides a  vortex-matter based picture of the transitions in the
most interesting part of the phase diagram, from a state with broken
$U(1) \times U(1)$ symmetry into a paired superfluid state and a
subsequent transition into a normal state. To find numerical backing
for the preemptive vortex-loop proliferation scenario, we perform a
large-scale Monte Carlo (MC) calculation of vortex matter in the
interacting BEC mixture using a representation in terms of the phases
of the ordering fields of the condensates. This numerical approach
allows us to study directly vortex matter
and therefore may be viewed as complementary to the
worm-algorithm based approach in
Refs. \onlinecite{amh1,amh2,DCP2}. The insight which we obtain from
Monte Carlo calculations on vortex matter may also shed light on how
the Andreev--Bashkin effect \cite{AB} modifies the vortex-matter phase
transition predicted for the liquid metallic state of hydrogen
\cite{LMH}.
\par
Finally, we remark that the problem of multicomponent vortex-loop
proliferation has a quite generic character, since it is also related
to a wide spectrum of phase transitions in other systems. An example
is represented by individually conserved  electrically charged condensates
that communicate with each other only via a fluctuating gauge field
\cite{GF,LMH,DCP1,DCP2,Balents_2005,KNS_2006,NKS_2007}. Moreover, a 
related problem arises in three-dimensional generalizations of phase 
transitions discussed recently for certain planar spin-$1$ condensates 
\cite{egor05,ashvin}.

\section{The model}
\par
We consider a mixture of Bose--Einstein condensates with $U(1)\times
U(1)$ symmetry and current-current interaction.  This system  in the
hydrodynamic limit is described by \cite{AB}
\begin{eqnarray}
  F&=& \frac{1}{2}\int_{\boldsymbol{r}}\du\bos{r}\left \{
    (\rho_1-\rho_d)\bos{v}_1^2
    +(\rho_2-\rho_d)\bos{v}_2^2
    +2\rho_d\bos{v}_1 \cdot \bos{v}_2
  \right \}\nonumber \\
  & = &
  \frac{1}{2}\int_{\boldsymbol{r}}\du\bos{r}\left \{
    \rho_1\bos{v}_1^2
    +\rho_2\bos{v}_2^2
    -\rho_d(\bos{v}_1 - \bos{v}_2)^2
  \right \},
\label{freeenergy} 
\end{eqnarray}
where $\bos{v}_i =\hbar \nabla \theta_i/m_i$.  The last term describes
a current-current interaction\cite{AB} (for its detailed microscopic
derivation, see Ref. \onlinecite{Fil}).  The microscopic origin of the
non-dissipative drag can for example be  elastic inter-component
scattering due to van der Waals forces between the charge-neutral
atoms in the system \cite{Fil}, or can also originate from a lattice 
\cite{amh1,Kuklov}.  This coupling is consistent with $U(1)\times U(1)$
symmetry and thus is  very different from the symmetry breaking intercomponent 
Josephson-coupling, which is a singular perturbation.  A drag term is perturbatively 
irrelevant and a critical strength is needed to change the zero-drag physics of the 
problem, due to the extra two gradients in the coupling between the two phases.
\par
The discrete model as such may also have a physical realization in
terms of a Bose--Einstein condensate on an optical lattice
\cite{amh1}. In the latter case, a particularly wide range of both
positive and negative $\rho_d$ can be accessed \cite{Kuklov}. The
parameter $\rho_d$ is a superfluid density of one condensate carried
by the superfluid velocity of the other as follows from the equations
of motion \cite{AB}, \beg
\bos{j}_1&=&(\rho_1-\rho_d)\bos{v}_1+\rho_d\bos{v}_2,  \\
\bos{j}_2&=&(\rho_2-\rho_d)\bos{v}_2+\rho_d\bos{v}_1.  \eee
Symmetry-restoring phase transitions in this system are associated
with proliferation of thermally excited topological defects, namely
vortex loops \cite{blowup}.  In what follows, we denote vortices in
the two-component condensate by a pair of integers corresponding to
the winding of the phases in each of the condensates
\beg
(\Delta\theta_1=2\pi n_1,\Delta\theta_2=2\pi
n_2)\equiv(n_1,n_2).
\eee
The current-current interaction $2\rho_d
\bos{v}_1 \cdot \bos{v}_2$ introduces a bias for 
counter-directed currents when $\rho_d$ is positive . Indeed, this term introduces an attractive
Coulomb interaction between $(\pm 1,0)$ and $(0,\mp 1)$ vortices. The
coefficients $\rho_1,\rho_2$ and $\rho_d$ must satisfy the relation
\begin{eqnarray}
  \rho_d < \frac{\rho_1\rho_2}{\rho_1+\rho_2}, 
\label{stability}
\end{eqnarray}
for stability. This puts an absolute upper bound on the amount of drag
in the system that can be considered physical. In the phase diagrams
to be presented below, we denote as gray (forbidden) those areas which
cover the sets of parameters that violate the above inequality.
\par

\section{\label{escales}Energy scales associated with bare
  stiffnesses}
Let us begin by a straightforward examination of the energy scales of
the problem. In what follows, we set the masses of the condensates
equal and absorb $\hbar/m$ in the definition of $\rho$. We focus on
the $\rho_d>0$ case. In what follows, we denote expressions for phase
stiffnesses for various topological excitations as $J_{(i,j)}$, (the
index $(i,j)$ refers to corresponding topological defect).  The
explicit expressions are given by \be J_{(1,0)}=\rho_1-\rho_d, \ee \be
J_{(0,1)}=\rho_2-\rho_d, \ee \be J_{(1,1)}=\rho_1+\rho_2, \ee \be
J_{(1,-1)}= \rho_1+\rho_2-4 \rho_d.  \ee
Let us denote the critical stiffness of the 3D$XY$-Villain model \cite{nguyen1998} by
\beg
\rho_c \approx 0.33
\eee
Then if we neglect any interactions between different species of vortices, naive 
estimates of the lines where various vortex modes would proliferate, are 
given as follows. 

$(1,0)$-vortices proliferate from an ordered background along a line
defined by $\rho_{1}-\rho_d=\rho_c$.

$(0,1)$-vortices proliferate from an ordered background along the
lines defined by $\rho_{2}-\rho_d=\rho_c$.

$(1,-1)$ vortices would proliferate from an ordered background along a
line defined by $ \rho_1+\rho_2-4\rho_d = \rho_c$.

Proliferation of $(1,1)$ is irrelevant because of the above types 
of topological excitations always proliferate (and thus restore 
symmetry) before $(1,1)$ vortices, when $\rho_d>0$.

Below we show that this naive energy-scale based picture is not 
correct.

\section{\label{PD}Phase diagram, equal
  stiffnesses \label{phase_diagram}}
The simplest case is where the bare phase stiffnesses of each
component is equal, so we begin by considering that case first.

 \subsection{\label{eqr1r2} Continuous phase transitions in limiting cases}
The character of the vortex-loop proliferation transition can readily
be understood in two limiting cases, by mapping the system to a single
component model yielding standard second order phase transitions.

One limit is the trivial limit $\rho_d \to 0$, when the system is
described by two independent $XY$ models undergoing a second order
phase transition from $U(1)\times U(1)$ to a symmetric state.  Indeed,
in this limit there is no energetic or entropic advantage in
restoring order by composite topological defects.

Another limit which is fairly simple to understand, follows from the
fact that by increasing $\rho_d$, the stiffness of $(1,-1)$-composite
defects can be made arbitrarily much smaller than the stiffnesses for $(1,0)$
and $(0,1)$ defects. This is the limit where $2\rho - 4\rho_d \approx
\rho_c < (\rho-\rho_d)$ and thus the vortex loop $(1,-1)$ costs little
energy to excite, while $(1,0)$ and $(0,1)$ effectively are frozen
out.  Physically, this also  means that in this limit it is energetically
costly to split a composite $(1,-1)$ defect into a pair of individual vortices, and 
therefore one may neglect its composite nature and map the system onto a $1$-component 
3D$XY$ model undergoing a phase transition at $J_{(1,-1)}=2\rho - 4\rho_d =
\rho_c$. { Because $(1,-1)$  vortices cannot disorder the phase sum,} this continuous 
phase transition is associated with going from a $U(1)\times U(1)$ state to a state 
with $U(1)$ symmetry associated with order in the phase sum, which is the 
``paired superfluid phase'' in Ref. \onlinecite{amh1}.
\par
Let us now consider the other regimes which occur in the case $\rho_1=\rho_2 =\rho$ case.
For some regimes another representation of Eq. (\ref{freeenergy}) will be useful, namely
\begin{multline}
    F= \frac{1}{2} \int_{\boldsymbol{r}}\du\bos{r} \Bigl\{ \left(
      \frac{\rho}{2} - \rho_d \right)
    \left[ \nabla(\theta_1-\theta_2) \right]^2  \\
    + \frac{\rho}{2} \left[ \nabla (\theta_1+ \theta_2)
    \right]^2 \Bigr\}. 
\label{freeenergy_sII}
\end{multline}
This form of the energy is particularly useful when we want to discuss
the  vortex matter of the remaining superfluid component in the
background of proliferated composite vortices.
We next proceed to
discussing this situation.

\subsection{\label{ptpreempr1eqr2} Phase transitions in a nontrivial
  vortex gas background}

A deviation from the vortex proliferation based on the naive energy
scales scenario is manifested in the transition to a fully symmetric
state in the regime $J_{(1,-1)} < J_{(1,0)}=J_{(0,1)}$, i.e.  $2\rho-
4\rho_d < \rho_c < \rho-\rho_d$. To understand how this takes place,
we should understand how the background of proliferated $(1,-1)$
vortices affects $(1,0)$ and $(0,1)$ vortices. This can be explained
from the separation of variables in Eq. \eqref{freeenergy_sII}. The
spontaneous proliferation of $(1,-1)$ vortices leaves the remaining
broken symmetry only in the second term. The corresponding remaining
phase stiffness is that of a ``clapping mode'' associated with a
response to varying the phase sum. The stiffness of the clapping mode
is destroyed by proliferation of the cheapest topological defects with
a winding in the phase sum. These defects are individual vortices
$(1,0)$ or $(0,1)$. The separation of variables
Eq. (\ref{freeenergy_sII}) suggests that the background of
proliferated $(1,-1)$ vortices destroys the phase stiffness in the
first term and thus only the second term determines the effective
stiffness of $(1,0)$ or $(0,1)$ vortices. Their stiffness is therefore
reduced compared to the bare stiffness in the naive energy scale
argument. The new effective stiffness is
$\tilde{J}_{(1,0)}=\tilde{J}_{(1,0)}=\rho/2$, and thus it suggests
that the system undergoes a phase transition to a fully symmetric
state at \be \frac{1}{2}\rho = \rho_c.
\label{fullysym}
\ee Note that from this argument, it follows that the proliferation of
$(1,0)$ or $(0,1)$ vortices in the background of proliferated $(1,-1)$
vortices is determined by $\rho$ only. This is testable in MC
calculations, and we report on it below.

\subsection{\label{preempr1eqr2}  Preemptive phase transition}
Now consider the most interesting regime where the line defined by the
relation $J_{(1,0)}=J_{(0,1)} = \rho-\rho_d=\rho_c$ intersects the
line defined by the relation $J_{(1,-1)} = 2\rho - 4\rho_d =
\rho_c$. We denote the intersection point derived from the naive energy scale-based
argument by $(\rho_I,\rho_{dI}) = (3 \rho_c/2,\rho_c/2)$.  Consider
the regime slightly above the point $\rho_I = 3 \rho_c/2$ (
  i.e.  $\rho=\rho_I+\delta$ and $\rho_d=\rho_{dI}+\delta/2$). Then,
from Eq. (\ref{freeenergy_sII}) we conclude that although the phase
transition is indeed initiated by proliferation of the
lowest-in-energy topological defects ($(1,-1)$ in this regime), the
remaining stiffness for $(1,0)$ and $(0,1)$ excitations $\rho/2
\approx \rho_I/2 = 3 \rho_c/4 $ (which can be read off from the second
term in Eq. (\ref{freeenergy_sII})), is actually less than
$\rho_c$. Hence, the vortices $(1,0)$ and $(0,1)$ cannot remain
confined once $(1,-1)$ are proliferated. Therefore, from the
separation of variables we may draw the conclusion that the simple
energy-scale based picture underestimates the critical
stiffnesses. More importantly, away from the limiting cases, the
process is cooperative and hence proliferation of composite defects
may trigger proliferation of individual vortices at a critical
stiffness where arguments based on energy scales alone would predict
that the individual vortex loops remain confined.  Thus, with respect
to $(1,0)$ and $(0,1)$ vortices, we are dealing with a ``preemptive''
vortex-loop proliferation scenario, triggered by the interaction with
vortices in a different sector of the model.  In the case where the
energy of $(1,-1)$ vortices is almost the same as that of $(1,0)$ and
$(0,1)$ vortices there is only one transition where by the same
arguments both types of topological defects assist each other in
restoring symmetry via a single phase transition. Numerical
calculations which we report in the second part of this paper confirm
this behavior of vortex matter. Importantly, whenever we observed this
behavior, the phase transition was first order within the resolution limits
of our MC calculations. The region of the
phase diagram showing first order transitions in our computations,  appears to be
consistent with the findings in the $J$-current
model\cite{amh1} { with the same symmetry}, though in our case the microscopic physics is
different. Note that this scenario is substantially different from the
continuous loop-proliferation transition invariably encountered in a
single-component model \cite{blowup,nguyen1998}.
\par
\begin{figure}[h!!]
  \includegraphics[width=\columnwidth]
  {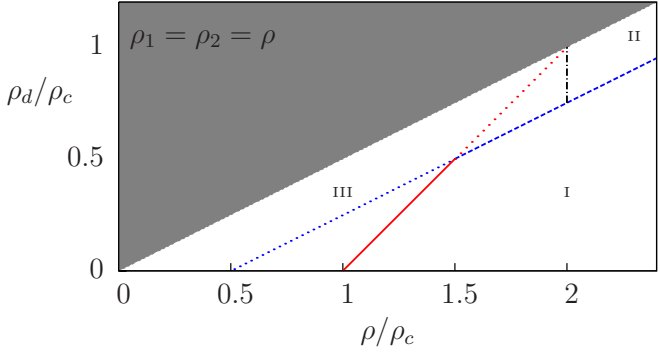}
  \caption{\label{PD1} (Color online) Schematic phase diagram which 
    follows 
    from  
    separations of variables
    for the model in Eqs. (\ref{freeenergy}) and
    (\ref{freeenergy_sII}), $\rho_1= \rho_2 =
    \rho$. The gray-shaded area is the forbidden 
    regime $\rho_d > \rho/2$. The lines separating the various regions
    are: a) {\underline{\it Solid (red) line}}:
    Proliferation of $(1,0)$ and $(0,1)$-vortices from an ordered
    background ({\it i.e.} large $\rho$), along the line $\rho_d= \rho-\rho_c$.  b)
    {\underline{\it Dashed (blue) line}}: Proliferation of
    $(1,-1)$-vortices from an ordered background, along the line
    $\rho_d= \rho/2-\rho_c/4$.  c) {\underline{\it Dashed-dotted
        vertical (black) line}}: Proliferation of $(1,0)$ and
    $(0,1)$-vortices from a background of proliferated
    $(1,-1)$-vortices, along the line $\rho = 2 \rho_c$.  This
    vortex-matter phase diagram has the same topology as that obtained
    from the $J$-current model \cite{amh1}. I:
    $U(1)\times U(1)$; II: $U(1)$-symmetry in the phase sum; and III:
    a fully symmetric case. }
\end{figure}

Fig. \ref{PD1} summarizes the new estimates for the lines of
vortex proliferation which follow from the separation of variable
argument Eq. (\ref{freeenergy_sII}).  They are given by three
different regimes.

$(1,0)$- and $(0,1)$-vortices proliferate from an ordered background
along a line defined by $\rho-\rho_d=\rho_c$ (solid red line in
Fig. \ref{PD1}),

$(1,-1)$ vortices proliferate from an ordered background along a line
defined by $2\rho - 4\rho_d= \rho_c$ (dashed blue line in
Fig. \ref{PD1}),

$(1,0)$- and $(0,1)$-vortices proliferate from a background of
proliferated $(1,-1)$-vortices at $\rho/2 = \rho_c$ (dashed-dotted black line in
Fig. \ref{PD1}).

\section{\label{uneqr1r2} Phase diagram, unequal stiffnesses}
We next generalize the above qualitative considerations to the case of
unequal stiffnesses $\rho_1 \neq \rho_2$. We will use the notation
that $\rho_2 = \alpha \rho_1; \rho_1 = \rho$.  So that the coefficient
$\alpha$ is a measure of the disparity of the stiffnesses. Since the
non-triviality of the phase diagram of this model is associated with
the possibility of tuning the energy of composite $(1,-1)$ defects to
be less than (or comparable to) the energy of individual vortices
$(1,0)$ and $(0,1)$, we use the same strategy as in the previous
section to analyze this model. That is, we need to separate the
variables by extracting all the stiffness terms which are
\emph{unaffected} by proliferation of $(1,-1)$ vortices. The
corresponding part of the free energy functional therefore should
depend on gradients of the phase sum only. Separating the variables in
such a way we arrive at the following representation of the model

\begin{multline}
      F =  \frac{1}{2} \int_{\bos{r}}\du\bos{r} \biggl\{ 
        \frac{\alpha \rho^2-(1+\alpha) \rho \rho_d}{(1+\alpha) \rho-4
          \rho_d}  \left[ \nabla\left(\theta_1 + \theta_2\right)
      \right]^2  \\
       \shoveright { + \frac{1}{(1 + \alpha) \rho-4 \rho_d} [ \left(\rho -2
        \rho_d\right) \nabla \theta_1 } - \left(\alpha \rho -2
        \rho_d\right) \nabla \theta_2 ]^2 \biggr\}.
    \label{freeenergy_sIIIb}
\end{multline}

Notice the asymmetric phase weights in the second term in contrast to
the symmetric separation of variables in Section \ref{PD}. The
asymmetry of the problem is also seen if we consider negative $\rho_d$
which would result in decreasing the energy of $(1,1)$ vortices
compared to $(1,0)$ and $(0,1)$ vortices.  Negative $\rho_d$ may
be easily realized in Bose-Einstein condensates on an optical lattice, and we
consider this possibility in Appendix B.
\par
In the following qualitative discussion in this section, we consider
only a positive $\rho_d$ and without loss of generality, we assume
that $\alpha > 1$. We start by going through the same energetics as we
did for the case $\rho_1=\rho_2$. First of all, the condition for
stability Eq. \eqref{stability}, now reads $\rho_d < \alpha
\rho/(1+\alpha)$. The proliferation of $(0,1)$-vortices from an
ordered background is now determined by the condition $\alpha
\rho-\rho_d = \rho_c$ or equivalently $\rho_d = \alpha \rho - \rho_c $
while that of $(1,0)$-vortices is determined by the condition $
\rho-\rho_d = \rho_c$, or equivalently $\rho_d = \rho-\rho_c$. These
lines now differ from each other, in contrast to the case $\rho_1 =
\rho_2$, and hence there will be one additional region in the phase
diagram. This follows, since at $\rho_d=0$, the phase transitions in
the model are expected to be two { non-degenerate} vortex-loop
proliferation transitions in the 3D$XY$-universality class, with a
regime with ordering only in one phase separating them. Moreover, the
spontaneous proliferation of $(1,-1)$-vortices from an ordered
background is now determined by the condition (to be read off from the
second term in Eq. (\ref{freeenergy_sIIIb})) $(1+\alpha)\rho- 4 \rho_d
= \rho_c$ or equivalently $\rho_d = [(1+\alpha) \rho -
\rho_c]/4$. These expressions reduce to those that were discussed in
Section \ref{preempr1eqr2} for the case $\alpha = 1$.
\par
Next, we proceed to investigate the condition for proliferation of
$(0,1)$- or $(1,0)-$vortices in a background of proliferated
$(1,-1)$-vortices. We thus assume (an assumption that will be checked
numerically in the second part of the paper) that the coefficient of
the stiffness associated with the second term in
Eq. (\ref{freeenergy_sIIIb}) has renormalized to zero. Then, the first
term accounts for the only phase stiffness remaining in the
system. Thus, the effective model becomes

  \begin{eqnarray}
    F^{{\rm eff}}_{(1,-1)} = \frac{1}{2}
    \int_{\bos{r}}\negthickspace \du\bos{r}
    \frac{\alpha \rho^2-(1+\alpha) \rho \rho_d}{(1+\alpha) \rho-4 \rho_d}
      \left[ 
      \nabla(\theta_1 + \theta_2) 
    \right]^2\negthickspace \negthickspace. \quad
    \label{freeenergy_sIIIc}
  \end{eqnarray}

\begin{figure}[h]
  \includegraphics[width=\columnwidth]
  {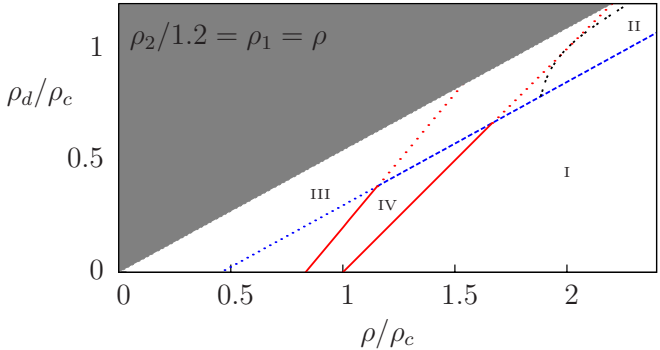}
  \caption{\label{PD2} (Color online) Schematic phase diagram with
    regions I,II, III, and IV for the model in Eqs. (\ref{freeenergy})
    and (\ref{freeenergy_sIIIb}) with $\rho_2= \alpha \rho_1 > \rho_1
    = \rho$. For the purposes of illustration, we have taken
    $\alpha=1.2$. The gray shaded are is the forbidden parameter
    regime $\rho_d > \alpha \rho/(1+\alpha)$. The lines separating the
    various regions are obtained as follows: a) {\underline{\it Solid
        (red) lines}}: Proliferation of $(0,1)$-vortices from an
    ordered background, along the line $\rho_d = \alpha \rho-\rho_c$,
    as well as proliferation of $(0,1)$-vortices from an ordered
    background, along the line $\rho_d = \rho-\rho_c$.  b)
    {\underline{\it Dashed (blue) line}}: Proliferation of
    $(1,-1)$-vortices from an ordered background, along the line
    $\rho_d = (1+\alpha)\rho/4-\rho_c/4$.  c) {\underline{\it
        Dashed-dotted  (black) line}}: Line of proliferation of individual vortices
    $(1,0)$ or $(0,1)$ in a background of proliferated vortices
    $(1,-1)$, given by Eq. \ref{prolif_line_alphagt1}. When we cross this line from 
    right to left, passing from region II to III, the
    stiffness of the clapping mode $\theta_1 + \theta_2$ is
    destroyed by the proliferation of individual vortices.  }
\end{figure}

Note the rather surprising fact that, provided the composite vortices
$(1,-1)$ have proliferated, the $(1,0)$- and $(0,1)$-vortices enter
the effective model on equal grounds even if the bare phase
stiffnesses for these differ.  The origin of this fact is that
$(1,-1)$ composite defects have an asymmetric effect on the partial
reduction of the bare phase stiffnesses of the individual vortices.
Thus $(1,0)$- or $(0,1)$ vortices will participate on equal grounds in
the restoration of the remaining symmetry.  Based on the above
conjectures we obtain the condition for proliferation of $(1,0)$- or
$(0,1)$ vortices in the background of proliferated $(1,-1)$ loops 
\begin{eqnarray}
 \frac{\alpha \rho^2-(1+\alpha) \rho \rho_d}{(1+\alpha) \rho - 4 \rho_d} = \rho_c. 
\end{eqnarray}
Observe that in contrast to the similar
condition Eq. (\ref{fullysym}) for the case of equal stiffnesses in
Section \ref{preempr1eqr2}, when $\alpha \neq 1$, $\rho_d$ no longer
drops out of this relation. The explicit relation is  
\begin{eqnarray}
\label{prolif_line_alphagt1} 
\rho_d = \frac{\alpha \rho^2 -(1+\alpha) \rho \rho_c}{(1+\alpha)\rho - 4 \rho_c}; \alpha \neq 1, 
\end{eqnarray} 
which is seen to approach
$\rho_d \to \alpha \rho/(1+\alpha)$ from below as $\rho$ becomes
large, i.e. the proliferation line approaches the forbidden
parameter region from below as $\rho \to \infty$.
The line of proliferation under discussion, namely the proliferation
of $(1,0)$- and $(0,1)$-vortices in the background of proliferated
$(1,-1)$-vortices, only comes into play above the
dashed  (blue) line separating phases I and II in Fig. \ref{PD2}. This is when 
the composite vortices are actually proliferated. We therefore only plot 
the line in this regime, and this is the dashed-dotted (black) line given 
in Fig. \ref{PD2}.
\par
In Fig. \ref{PD2}, the solid (red)
lines are the lines of proliferation of $(1,0)$- and $(0,1)$-vortices
from an ordered background. At $\rho_d = 0$, they emanate linearly
from $\rho = \rho_c/\alpha$ for $(1,0)$-vortices growing as $\alpha
\rho$, and from $\rho = \rho_c$ for $(0,1)$-vortices growing as
$\rho$.  The dashed (blue) line represents the line of proliferation
of $(1,-1)$-vortices from an ordered background. It emanates at
$\rho_d=0$ from $\rho = \rho_c/(1+\alpha)$, growing as $[ (1+\alpha)
\rho]/4$. The dashed-dotted (black) line represents the line across
which the effective stiffness of the clapping mode $\theta_1 +
\theta_2$ vanishes through the proliferation of individual vortices
$(1,0)$ or $(0,1)$. The lines are seen to divide the phase diagram
into four distinct regions, namely I) the completely ordered state,
II) the partially ordered state with proliferated $(1,-1)$-vortices
and confined individual $(1,0)$- and $(0,1)$-vortices, III) the
completely disordered state with proliferated individual vortices, and
IV) the partially ordered state with confined $(0,1)$-vortices and
proliferated $(1,0)$-vortices. Regions II and IV are therefore two
distinct partially ordered states with one broken $U(1)$-symmetry in
each case.

\subsection{\label{preempr1noteqr2} Preemptive scenario}
We next discuss the preemptive scenario for vortex-loop proliferation
for the more general case $\rho_1 \neq \rho_2$, largely following the
line of reasoning in Section \ref{preempr1eqr2}.  It turns out that
the physics is quite rich and markedly different from the
single-component case, which is rather surprising given the simplicity
of the coupling term between the two condensates,
cf. Eq. (\ref{freeenergy}). Hence, consider the intersection point
where the line of proliferation of $(1,-1)$-vortices from an ordered
background is intersected by the line of proliferation of
$(1,0)$-vortices from an ordered background.  This intersection takes
place at $(\rho_I,\rho_{dI}) = (3 \rho_c/(3-\alpha),\alpha
\rho_c/(3-\alpha))$.  Consider now a point slightly above the
intersection point above the line defined by the relation
$(\rho_I+\delta,\rho_{dI}+\delta_d) = (3
\rho_c/(3-\alpha)+\delta,\alpha\rho_c/(3-\alpha)+(1+\alpha)\delta/4)$,
where composite vortices are proliferated.  The remaining stiffness
for the clapping mode, $\rho_{\rm{clap}}$, is given by
Eq. (\ref{freeenergy_sIIIc})
\begin{eqnarray}
  \rho_{{\rm clap}} = 
      \frac{\alpha \rho^2-(1+\alpha) \rho \rho_d}{(1+\alpha) \rho-4 \rho_d}. 
  \label{rhoclap}
\end{eqnarray}
The question is now whether proliferation of $(1,-1)$ vortices can
trigger a preemptive proliferation of $(1,0)$ and $(0,1)$ vortices.
By evaluating $\rho_{\rm{clap}}(\rho,\rho_d,\alpha)$ at the
intersection point $(\rho_I, \rho_{dI})$ between proliferation of
individual vortices in an ordered background and proliferation of
composite vortices in an ordered background, the issue is if
$\rho_{\rm{clap}}(\rho_I,\rho_{dI},\alpha) <\rho_c$, ({\it i.e.} if
this estimate yields a situation that upon proliferation of composite
vortices the individual vortices no longer have enough stiffness
remaining to stay condensed).  If this is the case, then our estimates
will indicate a preemptive vortex-loop proliferation, following the
same line of reasoning as was used in Section \ref{preempr1eqr2} (to
be checked in Monte Carlo calculations in the second part of the
paper).

\section{Weighted phase sum order}
It has been observed in the past that in the drag problem
Eq. (\ref{freeenergy}), the vortices of the type $(1,-n)$ with $n>1$
can become energetically cheapest.\cite{Kuklov}. Let us apply the
separation of variables method to estimate analytically the position
and drag dependence of the transition lines in the phase diagram when
$(1,-n)$ -types of defects are relevant  as well as to describe how
vortex matter drives transitions from partially ordered to fully symmetric states
in these cases.    The accuracy of this method will be
checked numerically in the second part of the paper.
 
Consider $\rho_2<\rho_1$ and $\rho_d>0$.  First, one should examine
for which ratio of the bare stiffnesses $\rho_2/\rho_1$ does the
system prefer to proliferate composite $(1,-n-1)$ vortices rather then
$(1,-n)$.  The conditions when the energy for an $(1,-n-1)$ excitation
is less then that of an $(1,-n)$ excitation can be found as
follows. The phase stiffness associated with an $(1,-n)$
excitation is
$J_{(1,-n)}=\rho_1+n^2\rho_2-\left(1+n\right)^2\rho_d$. Hence one
finds that the inequality $\rho_1+(n+1)^2\rho_2 -(n+2)^2\rho_d <
\rho_1 +n^2 \rho_2 -(n+1)^2 \rho_d$ must be satisfied if the system is
to prefer proliferating $(1,-n-1)$ vortices in an ordered background
instead of proliferating $(1,-n)$ vortices. Combined with the
constraint $\rho_d<\rho_1\rho_2/(\rho_1+\rho_2)$ on $\rho_d$ we find
$\frac{2n+1}{2n+3}\rho_2<\rho_d<\frac{\rho_1\rho_2}{\rho_1+\rho_2}$
which gives
\begin{equation}
  \frac{\rho_2}{\rho_1}<\frac{1}{n+1/2}.
\end{equation}
This condition is illustrated in Table \ref{table_ratios}.  From this,
it follows that for $\rho_2/\rho_1<2/3$,  it is energetically less costly to excite
$(1,-2)$ vortices rather than $(1,-1)$ vortices for sufficiently large
value of $\rho_d$.
\begin{table}[h!]
  \begin{center}
    \caption{This table shows the condition for the ratio between the
      bare stiffnesses, when we assume that $\rho_2<\rho_1$ and
      $\rho_d>0$, for the system to proliferate a given composite
      vortex.}
    \label{table_ratios}
    \vspace{1ex}
    \begin{tabular*}{\columnwidth}{@{\extracolsep{\fill}}cccc}
      \hline
      &Composite vortex  & Condition&\\
      \hline
      &$(1,-1)$ &$2/3<\rho_2/\rho_1<1$&\\
      &$(1,-2)$ &$2/5<\rho_2/\rho_1<2/3$&\\
      &$(1,-3)$ &$2/7<\rho_2/\rho_1<2/5$&\\
      &\vdots&\vdots&\\
      \hline
    \end{tabular*}
  \end{center}
\end{table}

For such regimes the proper separation of variables is
\begin{multline}
  F=\frac{1}{2}{\int_{\bos{r}}}\du\bos{r} \biggl\{ \frac{\rho_1\rho_2
    -\rho_d\left(\rho_1+\rho_2\right)}
  {\rho_1+n^2\rho_2-\left(1+n\right)^2\rho_d}
  \left(n\nabla\theta_1 + \nabla\theta_2\right)^2 \\
  \shoveright{ \qquad \quad +
    \frac{1}{\rho_1+n^2\rho_2-\left(1+n\right)^2\rho_d} \biggl[
    \left(\rho_1-(1+n)\rho_d\right) \nabla\theta_1  } \\
  -\Bigl(\rho_2-(1+ \frac{1}{n}) \rho_d\Bigr) \nabla\theta_2 \biggr]^2
  \biggr\},
\end{multline}
were $n$ is an integer. This separation of variables is performed in
order to extract the part of the free energy which is unaffected by
$(1,-n)$ winding in the phases. Thus, upon proliferation of $(1,-n)$
vortices the system enters a phase with order in the weighted phase
sum $n\theta_1+\theta_2$ (while individual phases are disordered). The
effective phase stiffness which will remain in the system is given by

\begin{eqnarray}
  \negthickspace  \negthickspace  F^{{\rm
      eff}}_{(1,-n)}=\frac{1}{2}\int_{\bos{r}} \negthickspace
  \du\bos{r}
  \frac{\rho_1\rho_2 -\rho_d\left(\rho_1+\rho_2\right)}
  {\rho_1+n^2\rho_2-\left(1+n\right)^2\rho_d}
  \left(n\nabla\theta_1 + \nabla\theta_2\right)^2 \negthickspace
  \negthickspace . \quad
  \label{1-n}
\end{eqnarray}

\begin{figure}[h]
  \includegraphics[width=\columnwidth] {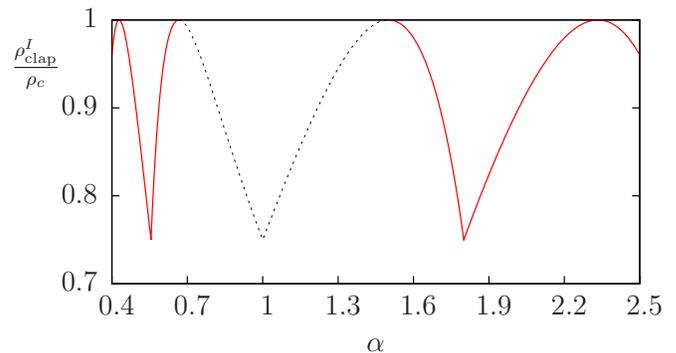}
  \caption{\label{rho_clap_alpha} (Color online) Plot of
    $\rho_{\rm{clap}}(\rho_I,\rho_{dI},\alpha)/\rho_c$ as a function
    of $\alpha$. Note that the system is symmetric around
    $\alpha=1$. The dashed (black) line in the above figure shows that
    the remaining stiffness of the clapping mode (in the background of
    proliferated composite $(1,-1)$-vortices) is less than the
    critical coupling for vortex-loop proliferation in a parameter
    regime $2/3 < \alpha < 3/2$. This is the parameter regime where it
    is correct to limit oneself to the sector where the composite
    proliferated background vortices are of the type $(1,-1)$.  For
    $\alpha < 2/3$, the composite proliferated background vortices are
    of type $(-n,1)$, while for $\alpha > 3/2$, the composite
    proliferated background vortices are of type $(1,-n)$.  The fact
    that $\rho_{\rm{clap}}(\rho_I,\rho_{dI},\alpha) < \rho_c$
    indicates that we are in a parameter regime where the full
    restoration of $U(1) \times U(1)$-symmetry proceeds from a
    preemptive
    vortex-loop proliferation phase transition, as explained in the
    text. The solid (red) line in the above figure shows
    $\rho_{\rm{clap}}(\rho_I,\rho_{dI},\alpha)$ for $\alpha < 2/3$ in
    a regime where $(2,-1)$-vortices  trigger preemptive
    proliferation of all topological defects and $\alpha > 3/2$ where $(1,-2)$-vortices
    initiate the phase transition into fully symmetric state.}
\end{figure}

In contrast to the case considered in previous sections,  here the
individual phases do not participate on equal grounds after
proliferation of $(1,-n)$ vortices  because
one of the phases has a factor $n$ and is therefore more expensive to fluctuate.
Nonetheless,  there are several types of
topological defects which can contribute on equal grounds to restore
the remaining symmetry.
 In Fig. \ref{rho_clap_alpha} we plot
$\rho_{\rm{clap}}(\rho_I,\rho_{dI},\alpha)/\rho_c$ as a function of
$\alpha$.
\par
For definiteness, we next consider in detail the case $n=2$. Then, the cheapest topological 
defect with which to restore the symmetry in Eq. (\ref{1-n}) is given a doublet of an 
elementary vortex $(0,1)$ and a  composite vortex $(1,-1)$ which is of lower order than the 
vortex $(1,-2)$ which drives the system into the partially ordered state  (\ref{1-n}). The 
transition back to a fully symmetric state then takes place when 
\begin{eqnarray}
  \frac{\rho_1\rho_2 -\rho_d\left(\rho_1+\rho_2\right)}
  {\rho_1+n^2\rho_2-\left(1+n\right)^2\rho_d}
  = \rho_c.
  \label{rhoc}
\end{eqnarray}
In our MC calculations, which we report below, we check this dependence.

Before we proceed to the Monte Carlo calculations, we remark on the accuracy of the 
estimates of the location of the phase-transition lines based on separation of variables. 
The location of the phase-transition lines based on the above arguments, have corrections 
in the regimes of the phase diagram where several such lines split. This is because in the 
vicinity of such splitting points, the  energy scales associated with various types of 
topological defects are not well separated. Hence, energetically next-to-cheapest excitations 
could participate in the depletion of the phase stiffness. 
The above arguments become more 
accurate as we move away from splitting points. However, they underestimate critical stiffnesses 
near splitting points. Below, we  perform Monte Carlo simulations to study the least analytically 
tractable region near the splitting points. We find that even near the splitting points, the 
separation-of-variables based argument is quite accurate.

\section{\label{MC_results} Monte Carlo calculations}

We next proceed to presenting our numerical results based on large-scale
Monte Carlo calculations, for which we need to define our continuum
model on a numerical lattice. Alternatively, we may view it as a
physical realization of a $2$-component Bose--Einstein condensate on
an optical lattice, as alluded to above. Providing a faithful lattice
representation of the continuum model Eq. (\ref{freeenergy}) using
phase variables is not straightforward, as some of the schemes for
formulating the theory on a lattice, which are standard in the
single-component case, introduce subtle artifacts when the
current-current interaction between two condensates is discretize. It
turns out that a study of the vortex physics in a lattice
representation of the model Eq. (\ref{freeenergy}) is best facilitated
by the so-called Villain approximation. This accommodates the
compactness of the superfluid phase of the ordering fields and
accounts properly for the current-current interaction. The Villain
Hamiltonian for the two-component condensate is given by
\begin{widetext}
  \begin{eqnarray}
    \label{eq:VillainH}
    H_v\left[\bos{\Delta}\theta_1,\bos{\Delta}\theta_2\right] 
    & = &
    \sum_{\bos{r},\mu}V_\mu\left(\Delta_\mu\theta_1,\Delta_\mu\theta_2;T
    \right),
    \nonumber \\ 
    V_\mu(\chi_1,\chi_2;T) & = &
    -\beta^{-1}\ln\biggl\{\negthickspace \negthickspace \sum_{\quad
      n_{1,\mu},n_{2,\mu}}\ex{-\beta/2  
      \left[\rho_1(\chi_1-2\pi n_{1,\mu})^2 + \rho_2(\chi_2 -2\pi
        n_{2,\mu})^2 -\rho_d(\chi_1-\chi_2 -2\pi(n_{1,\mu}-n_{2,\mu}))^2
      \right]} \bigg\}, 
  \end{eqnarray}
\end{widetext}
where the partition function of the system is given by
$Z=\int_{0}^{2\pi} \mathcal{D}\theta_1\mathcal{D}\theta_2\ex{-\beta
  H_v}$, and $\beta=1/k_B T$.  We have performed Monte Carlo
calculations on Eq. \eqref{eq:VillainH}, using local Metropolis
updating of the fields, $\theta_1(\bos{r})$,$\theta_2(\bos{r})\in
[0,2\pi)$, while ensuring that $\bos{\Delta}\theta_i(\bos{r})\in
[-\pi,\pi)$. The system sizes considered were $L\times L\times L$ with
$L= 16, 24, 32, 40, 48, 56$ and $64$. We have chosen $\beta = 1$ and
varied $\rho = \rho_1 = \rho_2/\alpha$. Additionally, the drag
$\rho_d$ is chosen proportional to $\rho$, and thus, there is
technically no difference between this approach and varying the
temperature for fixed $\rho,\rho_d$.  During the computations, we
sample the total energy $H_v$ of the system, and various helicity
moduli. There are six different helicity moduli we keep track of (not
all independent). The most general helicity modulus one can define in
this system is applying a twist $\theta_1 \to \theta_1
+a_1\bos{r}\cdot\hat{\mathrm{e}}_\mu \delta$ and $\theta_2 \to
\theta_2 + a_2\bos{r}\cdot\hat{\mathrm{e}}_\mu\delta$. The helicity
modulus is then given as the second derivative of the free energy with
respect to $\delta$. For details, see Appendix \ref{helicity}.
We measure the helicity modulus associated with six different choices
of twists, $(a_1=1, a_2=0)$, $(a_1=0, a_2=1)$, $(a_1=1,a_2=\pm 1)$ and
$(a_1=1,a_2=\pm 2)$ {\it i.e.} twists in $\theta_1$, $\theta_2$,
$\theta_1\pm\theta_2$ and $\theta_1\pm 2\theta_2$, respectively. These
are denoted $\Upsilon_1^{\mu}$, $\Upsilon^{\mu}_{2}$,
$\Upsilon^{\mu}_{\pm}$ and $\Upsilon^{\mu}_{1,\pm 2}$. Here,
$\Upsilon^{\mu}_{\pm} = \Upsilon^{\mu}_{1} \pm 2\Upsilon^{\mu}_{12} +
\Upsilon^{\mu}_{2}$ and $\Upsilon^\mu_{1,\pm 2}=\Upsilon^\mu_1 \pm
4\Upsilon^\mu_{12} + 4\Upsilon^\mu_2$.  A finite helicity modulus is a
signal of a finite superfluid density of the associated quantity, a
finite $\Upsilon^\mu_{\pm}$ represents the possibility of having
co-(counter-)superflow of the two components.  Likewise, the vanishing
of the helicity moduli $\Upsilon_{a_1,a_2}^{\mu}$ signals a thermally
driven spontaneous proliferation (blowout) of vortex loops originating
with multiples of $2 \pi$-windings in the phases $a_1\theta_1 +
a_2\theta_2$. We have considered these quantities for equal as well as
for different bare phase stiffnesses $\rho_1$ and $\rho_2$, and have
in all cases varied the drag coefficient $\rho_d$ from $0$ up to the
maximum allowed value compatible with the stability of the
two-component superfluid ground state.  The location of the phase
transitions are read off from the peak in the heat capacity
\par
\begin{figure}[h]
  \includegraphics[width=\columnwidth]{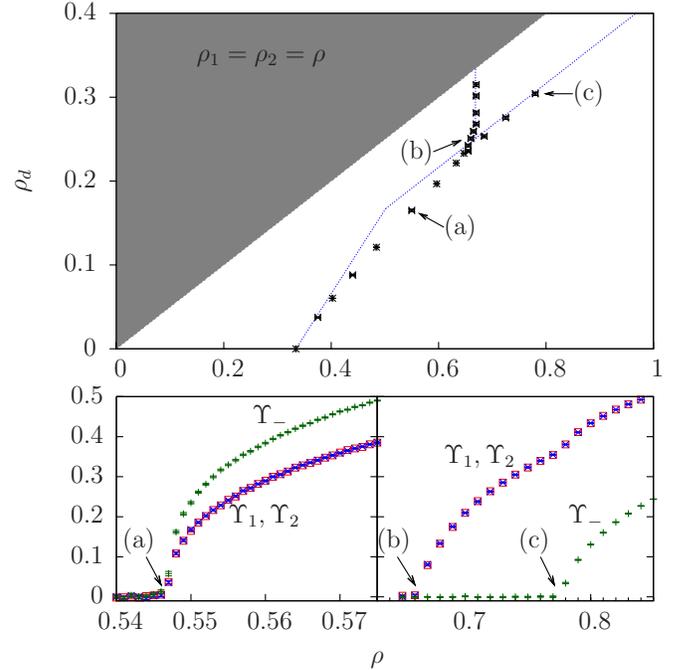}
  \caption{\label{PDHel} (Color online) Phase diagram and a set of
    helicity moduli for the model Eq. (\ref{eq:VillainH}) with equal
    bare stiffnesses . The shaded region illustrates the forbidden
    parameter regime $\rho_d > \rho/2$.  The helicity moduli are
    $\Upsilon_1$,$\Upsilon_2$, and $\Upsilon_{-}$.  The leftmost
    helicity moduli are measured for a drag $\rho_d=0.30\rho$, while
    the rightmost for $\rho_d=0.39\rho$.  }
\end{figure}

We first discuss the case $\rho_1=\rho_2$, for which results for the
phase diagram and helicity moduli are shown in Fig. \ref{PDHel}. The
dotted lines represent the predictions based on our  analytical arguments
from the previous sections.
At $\rho_d=0$, the system features a doubly degenerate phase
transition from a $2$-component superfluid to a $2$-component normal
fluid at the critical couplings $\rho_{c1} = \rho_{c2} = 0.33$. These
phase transitions are in the 3D$XY$-universality class. When drag is
introduced, it initially has the effect of reducing the stiffnesses of
the individual phases $\theta_1$ and $\theta_2$, thus moving the
doubly degenerate phase transitions to higher couplings
$(\rho_{1c},\rho_{2c})$. At large enough drag these phase transitions
split, and the intermediate phase with ordering only in the phase sum
emerges (the ``paired superfluid phase'' in terms of
Ref. \onlinecite{amh1}). We observe that our computations show that
the analytic arguments advanced in previous sections describe quite
accurately the phase diagram.

The line of transition from $U(1)\times U(1)$ to a fully symmetric
phase changes its slope indicating that composite vortices for sufficiently large drag 
initiate the transition into the symmetric state (the preemptive vortex-loop proliferation 
scenario). Importantly, near the bending point the  actual transition line is situated to 
the right of the dotted lines, which originate with the above bare-stiffness arguments  
when sub-leading type of topological defects are not taken into account. Therefore,  these 
estimates naturally underestimate the stiffness at the actual position of a preemptive 
transition. However, even in this region, the deviation is not significant.


The transition line from the state with ordering only in the phase sum
to a fully symmetric state precisely coincides with the analytic
estimates and is independent of $\rho_d$, in the equal stiffnesses
case, away from the splitting point.  The splitting point takes place
at significantly higher coupling constants 
in the phase diagram than what the naive energy-scale based argument gives, 
and is also in  good agreement with the splitting point of the preemptive 
loops proliferation scenario discussed in Sections \ref{preempr1eqr2} and 
\ref{preempr1noteqr2}.
\par
The corresponding results for the various helicity moduli are also
shown in Fig. \ref{PDHel}. In the lower right panel the helicity
modulus $\Upsilon_{-}$ for the composite vortex mode $(1,-1)$ vanishes
first as we approach lower couplings (or equivalently, higher
temperatures) from the completely ordered side. The resulting state is
only partially ordered.  The individual stiffnesses $\Upsilon_1$ and
$\Upsilon_2$ vanish simultaneously at some lower coupling (higher
temperature), rendering the system a normal fluid. The interesting
part of the phase diagram is just below the splitting point, where we
have a region in which the phase transition is first-order.
\begin{center}
  \begin{figure}[h]
    \includegraphics[width=\columnwidth] {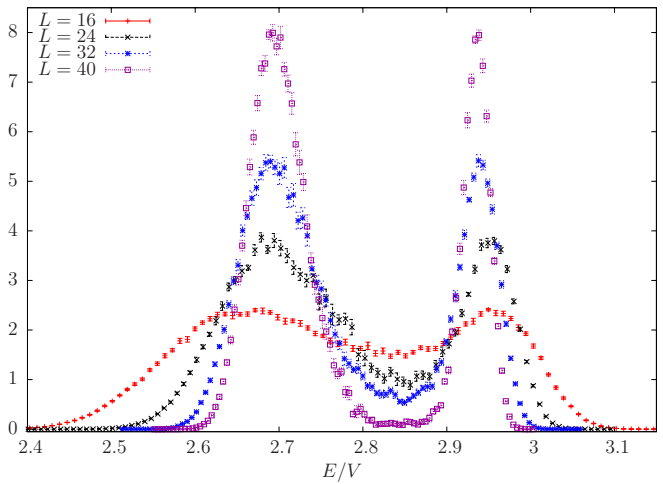}
    \caption{\label{HistI} (Color online) The energy histograms for
      $(\rho,\rho_d) \approx (0.60,0.20)$ , with $\alpha=1$, {\it i.e.}
      in the preemptive region. A clear double peak structure is seen
      to develop, an indication of a first order transition. The areas
      under the histograms are normalized to 1.  }
  \end{figure}      
\end{center}         

We find strong indications, shown in Fig. \ref{HistI}, that the
transition from the $U(1)\times U(1)$ state to the fully symmetric
state in the region where vortex-matter based argument suggest preemptive
scenario is indeed a first order
transition. This is also in agreement with previous computations of
the $J$-current model  \cite{amh1}.

\begin{figure}[h]
  \includegraphics[width=\columnwidth]
  {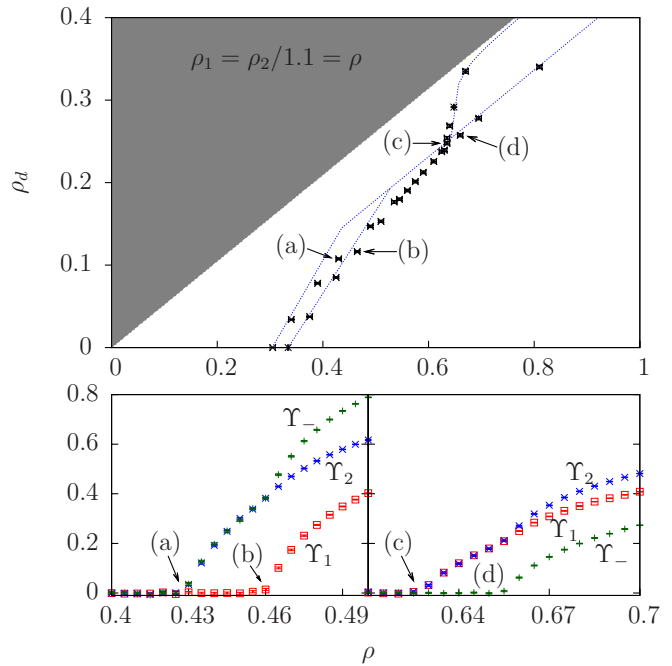}
  \caption{\label{PDHelII} (Color online) Phase diagram and a set of
    helicity moduli for the model Eq. (\ref{eq:VillainH}), for
    $\alpha=1.1$. The shaded region illustrates the forbidden
    parameter regime $\rho_d > \rho_1 \rho_2/(\rho_1+\rho_2)$. The
    helicity moduli are $\Upsilon_1$,$\Upsilon_2$, and $\Upsilon_{-}$.
    The left most helicity moduli are measured for a drag
    $\rho_d=0.25\rho$, while the rightmost are $\rho_d=0.39\rho$.}
\end{figure}

\begin{center}
  \begin{figure}[h]
    \includegraphics[width=\columnwidth] {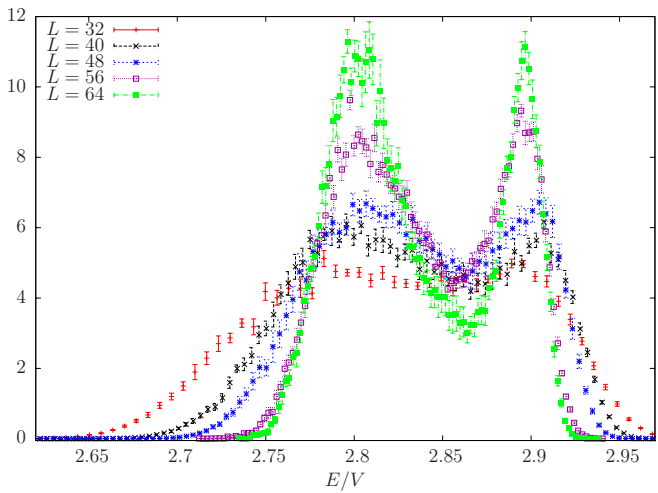}
    \caption{\label{HistII} (Color online) The energy histograms for
      $(\rho, \rho_d) \approx (0.60,0.22)$, with $\alpha=1.1$, i.e. in the
      preemptive region.  A clear double peak structure is seen to
      develop, an indication of a first order phase transition. The
      areas under the histograms are normalized to 1.}
\end{figure}
\end{center}

\begin{center}
  \begin{figure}[h]
    \includegraphics[width=\columnwidth] {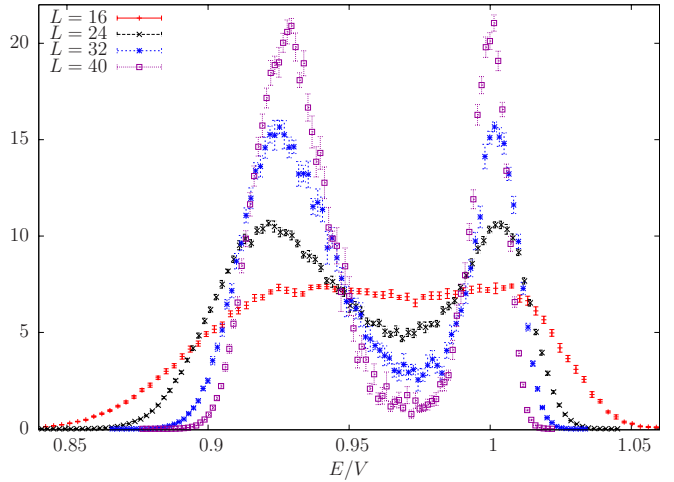}
    \caption{\label{HistIII} (Color online) The energy histograms for
      $(\rho, \rho_d) \approx (1.77,0.58)$, with $\alpha=0.55$, i.e. in the
      preemptive region. A clear double peak structure is seen to
      develop, an indication of a first order transition. The areas
      under the histograms are normalized to 1.  }
  \end{figure}
\end{center}

We proceed to discuss the case of slightly unequal bare stiffnesses,
{\it i.e.} $(1,-n)$ vortices with $n>1$ are unimportant. In our
computations, we have used $\rho_2=1.1\rho_1$, see
Fig. \ref{PDHelII}. At $\rho_d=0$ the system features two independent
phase transitions in the 3D$XY$-universality class at
$\rho_{c1}\approx 0.33$ and $\rho_{c2}\approx 0.30$. When drag is
introduced, it initially has the effect of driving the transitions to
higher values of $\rho$ (lower values of $T$). For moderate values of
drag, these two transition close in on each other, before they merge
into one transition from a $U(1)\times U(1)$ state into the symmetric
state.  In terms of vortex matter, this is the preemptive region of the phase diagram. For even
larger drag this line splits, and the intermediate phase with ordering
associated with the phase sum emerges.
The dotted lines in Fig. \ref{PDHelII} are predictions described in
Section \ref{uneqr1r2} and these agree well with our
computations. Specifically, we observe that when the helicity modulus
$\Upsilon_-$ is renormalized to zero, the individual stiffnesses
become equal, as expected from our separation of variables arguments
Eq. \eqref{freeenergy_sIIIc}. Moreover, in Fig. \ref{HistII}, we show
the corresponding energy histograms computed on the phase-transition
line between points (b) and (c) in Fig. \ref{PDHelII}, namely at
$(\rho, \rho_d) \approx (0.60,0.22)$. This puts us in a part of the phase diagram where we would
expect, based on our  vortex-matter arguments, to be able to see the
preemptive scenario explained above played out. Indeed, the phase
transition is clearly seen to be of first order also in this case,
thus confirming that the preemptive  vortex-loop proliferation scenario is also realized for
unequal bare phase stiffnesses $\rho_1$ and $\rho_2$.
We now discuss the case of significantly different bare stiffnesses,
i.e. when $(1,-n)$-vortices with $ n > 1 $ are important. In our
computation we have used $\rho_2=0.55\rho_1$, which from Table
\ref{table_ratios} indicate that we should observe a state with order
in the weighted phase sum, with $n=2$.
\begin{figure}[h]
  \includegraphics[width=\columnwidth]
  {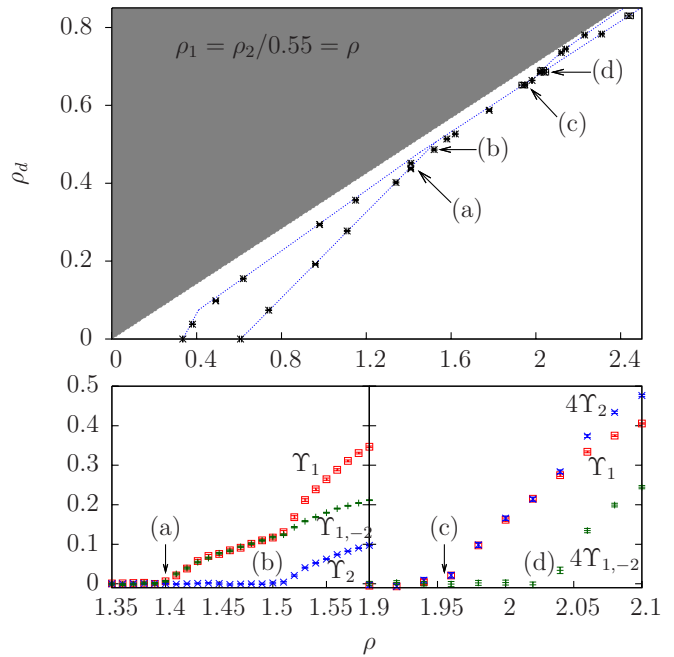}
  \caption{\label{PDHelIII} (Color online) Phase diagram and a set of
    helicity moduli for the model Eq. (\ref{eq:VillainH}),
    $\alpha=0.55$. The shaded region illustrates the forbidden
    parameter regime $\rho_d > \rho_1 \rho_2/(\rho_1+\rho_2)$. The
    helicity moduli are $\Upsilon_1$,$\Upsilon_2$, and
    $\Upsilon_{1,-2}$. Here, the latter correspond to $\Upsilon_{-}$,
    with the difference that the $\theta_2$-phase is twisted twice as
    much as $\theta_1$. The leftmost helicity moduli are measured for
    a drag $\rho_d=0.32\rho$, while the rightmost are
    $\rho_d=0.336\rho$.  }
\end{figure}
As in the case of slightly unequal stiffnesses the system features two
independent transition in the 3D$XY$-universality class, in our
computation the transitions at $\rho_d = 0$ occurs at
$\rho_{c1}\approx 0.33$ and $\rho_{c2}\approx 0.605$. At small drag
values the transitions stay independent and are shifted to higher
values of $\rho$. For moderate drag values the region with partial
order (order in $\theta_1$) becomes smaller, before disappearing at
some higher drag value. The system then enters into the preemptive
 vortex-loop proliferation region, where the system features a transition from a $U(1)\times
U(1)$-state to the fully symmetric state. In this region we find
strong indications of first order transitions, shown in
Fig. \ref{HistIII}.
For even larger drag values this single lines splits into two lines,
and a partially ordered state appears. The partially ordered state
which appears is a state were $(1,-2)$-vortices have proliferated
while individual vortices stay confined (in general $(1,-n)$-vortices  can
proliferate).  In the lower rightmost panel in Fig. \ref{PDHelIII} we
observe that $\Upsilon_{1,-2}$ drops to zero while $\Upsilon_1$ and
$\Upsilon_2$ remain finite, from this we conclude that we observe the
partially ordered state with order in the weighted phase sum. 
{The MC calculation shows that in this quite generic case
that the analytical arguments from the first part of the paper
are remarkably quantitatively accurate even near line-splitting points.
Observe
further that when $\Upsilon_{1,-2}$ have renormalized to zero we have
the relation $\Upsilon_1 = 4\Upsilon_2$, as expected from the discussion near}
Eq. \eqref{1-n} (in general we expect $\Upsilon_1 = n^2\Upsilon_2$).

\section{Implication for the liquid metallic hydrogen problem}
The approach developed above is also useful to obtain  insight into 
the role of non-dissipative drag when it is included in the problem of 
multicomponent electrically charged condensates \cite{LMH,GF}. For example, 
in the problem of projected quantum fluid states of hydrogen, we deal with 
two electrically charged fields corresponding to electronic and protonic
condensates. The fields will then be coupled by an electromagnetic gauge 
field in addition to the now familiar drag coupling
\begin{align}
F=\frac{1}{2}\int_{\bos{r}}\du\bos{r}
\big\{
&\rho\left(\nabla\theta_1-e\bos{A}\right)^2
+\alpha\rho\left(\nabla\theta_2+e\bos{A}\right)^2\nonumber \\
-&\rho_d\left(\nabla\theta_1-\nabla\theta_2-2e\bos{A}\right)^2
+\left(\nabla\times\bos{A}\right)^2
\big\},
\end{align}
were $e$ is the charge and $\bos{A}$ is the gauge field.
\par
In this case, the physically relevant separation of variables corresponds to
extraction of the phase sum. This follows, since in this situation it is the
phase \emph{sum} which is not coupled to the gauge field. This allows us to
draw conclusions about superfluid and superconducting states of the system 
\cite{LMH}. Following the same line of reasoning, an assessment of the role 
of non-dissipative drag is made by an extraction of the phase sum to distinguish 
the drastically different charged and neutral modes of the system. The model 
then becomes
\begin{multline}
  F = \frac{1}{2} \int_{\bos{r}}\du\bos{r} \biggl\{ 
    \frac{\alpha \rho^2-(1+\alpha) \rho \rho_d}{(1+\alpha) \rho-4
      \rho_d}  \bigl[ \nabla\left(\theta_1 + \theta_2\right)
  \bigr]^2  \\
  \shoveright{ + \frac{1}{(1 + \alpha) \rho-4 \rho_d} \bigl[ \left(\rho -2
      \rho_d\right) \nabla \theta_1 -\left(\alpha \rho -2
      \rho_d\right) \nabla \theta_2 } \\
  -e\{(1+\alpha)\rho -4\rho_d \} \bos{A} \bigr]^2
  +(\nabla\times\bos{A})^2\biggr\}.
    \label{LMH}
\end{multline}
By virtue of featuring one composite charged mode and one
composite neutral mode, this model has the same structure as
the model with zero drag \cite{LMH,GF}. However, now the stiffnesses of neutral
and charged modes acquire dependence to the drag coefficient $\rho_d$. Therefore, the 
conclusions of  Ref. \onlinecite{LMH} should be rather robust against finite-drag perturbations. 
The inter-component drag term is a quite different perturbation to the system 
compared to inter-component Josephson coupling. The latter is prohibited in 
hydrogen, but is allowed in multicomponent electronic condensates. Josephson 
coupling amounts to an explicit symmetry breakdown, and in terms of long length
scale physics it represents a singular perturbation compared to the case where 
it is absent. An inter-component drag term has two gradients in it, since it 
is a current-current interaction. Consequently, it has a naive scaling dimension 
which is reduced by $2$ compared to the Josephson coupling, and in contrast to
the Josephson couplling it does not represent a singular perturbation. Quite the 
contrary, as we have seen, a critical value of the strength of the inter-component 
drag term is required for it to have an appreciable effect on the physics of the 
system. 

\section{Summary and conclusions}
In this paper, we have studied the problem of the influence of
non-dissipative inter-component drag on the phase diagram and phase
transitions in a two-component Bose--Einstein condensate. The
non-dissipative drag is a quite generic feature present in interacting
multicomponent systems in the continuum as well as on a lattice
\cite{AB,AB-2,Kuklov,Fil}. Recently, the topology of the phase diagram and
orders of the phase transitions were intensively studied in
the $J$-current model with $U(1)\times U(1)$ symmetry by means of
worm-algorithm based Monte Carlo simulations \cite{amh1,Kuklov1,Kuklov,amh2},
revealing novel features such as conversions of the phase transitions
from continuous to first order as a function of drag strength.
\par
We have developed an approach in terms of topological defects for
understanding these phase transitions and get new insight into physics
of the various states of two-component Bose--Einstein condensates. We
have carried out an investigation of the phase diagram based on
analytical vortex-matter arguments, and suggested a novel scenario of
vortex-matter behavior, namely a ``preemptive vortex-loop
proliferation''.  
Such a scenario may well be generic to systems where
symmetry is restored through proliferation of distinct topological
defects in the form of vortex loops that have been excited out of the
individually conserved condensates. We have found support for these
scenarios in large-scale Monte Carlo calculations. These computations
have been carried out using a representation of the system in terms of
the phase of the complex ordering field of each of the
components. The approach allows us to investigate directly the
physics of topological defects in this system. Importantly, the phase
representation also allows us to study the system under rotation. This
can provide a bridge for studying these states of matter
experimentally via rotational response. Work on this problem is in
progress \cite{eskil_rotation}.

\acknowledgements{The
  authors acknowledge useful discussions with  J. Hove, D. A. Huse,  
  A. Kuklov, E.J. Mueller B. Svistunov,
  and M. Wallin. This work was supported by the Norwegian Research
  Council Grant Nos. 1585187/431, 158547/431 (NANOMAT), and Grant
  No. 167498/V30 (STORFORSK). The authors thank the Center for
  Advanced Study at the Norwegian Academy of Science and Letters,
  where part of this work was done.}

\appendix

\section{\label{helicity} Superfluid density matrix in a $2$-component
  system}

In general, the helicity modulus defines the superfluid density of a
system. For the {Andreev--Bashkin problem \cite{AB}}, the superfluid
density is a matrix quantity given by the second derivative of the
free energy of the system with respect to an infinitesimal twist in
the phase, i.e.  $\theta(\bos{r})\rightarrow
\theta(\bos{r})-\bos{\delta}\cdot\bos{r}$. The helicity modulus,
$\bos{\Upsilon}$, is then given as $\Upsilon_{\mu} =
\left.\frac{1}{L^3}\frac{\partial^2 F[\bos{\delta}]}{\partial
    \delta_\mu \partial \delta_\mu}\right|_{\delta=0}$. Since
$F[\bos{\delta}]=-\beta^{-1}\ln Z[\bos{\delta}]$, where $\beta$ is
inverse temperature and $Z[\bos{\delta}] = \int \mathcal{D}\Gamma
\ex{-\beta H[\bos{\delta}]}$ is the partition function, the helicity
modulus can further be written as
\begin{equation}
  \begin{split}
    \Upsilon_\mu = \frac{1}{L^3} \Biggl[ &
    \avg{\frac{\partial^2 H[\bos{\delta}]}{\partial \delta_\mu^2}}\\
    & -\beta \avg{ \left( \frac{\partial H[\bos{\delta}]}{\partial
          \delta_\mu} -\avg{ \frac{\partial H[\bos{\delta}]}{\partial
            \delta_\mu} } \right)^2 } \Biggr] \Biggr|_{\delta=0}.
      \end{split}
\end{equation}
This is a general expression for the helicity modulus, independent of
the form of the Hamiltonian.  We now specify the form of the
Hamiltonian to that of a two-component Villain-model i.e.
$H_v=\sum_{\bos{r},\nu}V_\nu(\Delta_\nu\theta_1(\bos{r}),\Delta_\nu\theta_2(\bos{r}))$
where the potential $V_\nu$ is given in Eq.(\ref{eq:VillainH}). We now
apply an arbitrary twist in the phases,
$\left(\newatop{\theta_1(\bos{r})}{\theta_2(\bos{r})}\right) \to
\left(\newatop{\theta_1(\bos{r})}{\theta_2(\bos{r})}\right) -
\left(\newatop{a_1}{a_2}\right)\bos{\delta}\cdot\bos{r} $, were
$a_1,a_2$ are two real numbers and expressions on both side of the
arrow satisfies periodic boundary conditions. The Hamiltonian then
takes the form
$H_v[\bos{\delta}]=\sum_{\bos{r},\nu}V_\nu(\Delta_\nu\theta_1(\bos{r})-a_1\delta_\nu,\Delta_\nu\theta_2
(\bos{r})-a_2\delta_\nu)$.  The first and second derivatives of the
Hamiltonian are then given by,
\begin{widetext}
  \begin{align}
    \left.\frac{\partial H[\bos{\delta}]}{\partial
        \delta_\mu}\right|_{\bos{\delta}=0}&=
    \sum_{\bos{r}}\left(-a_1\frac{\partial V_\mu}{\partial
        \Delta_\mu\theta_1}-a_2\frac{\partial V_\mu}{\partial
        \Delta_\mu\theta_2} \right)\\
    \left.\frac{\partial^2 H[\bos{\delta}]}{\partial
        \delta_\mu^2}\right|_{\bos{\delta}=0} &=
    \sum_{\bos{r}}\left(a_1^2\frac{\partial^2
        V_\mu}{\partial\Delta_\mu\theta_1^2} +
      2a_1a_2\frac{\partial^2V_\mu}{\partial\Delta_\mu\theta_1\partial\Delta_\mu\theta_2}
      +a_2^2\frac{\partial^2 V_\mu}{\partial\Delta_\mu\theta_2^2}
    \right).
  \end{align}
  The helicity modulus associated with this choice of twist in the
  phase, is given by
  \begin{align}
    \Upsilon^\mu_{a_1,a_2}= &\frac{a_1^2}{L^3} \left[
      \avg{\frac{\partial^2H_v}{\partial\Delta_\mu\theta_1^2}}
      -\beta\avg{\left(\frac{\partial
            H_v}{\partial\Delta_\mu\theta_1}-\avg{\frac{\partial
              H_v}{\partial\Delta_\mu\theta_1}}\right)^2}
    \right]\label{hel_mod1}\\
    & +\frac{2a_1a_2}{L^3}\left[
      \avg{\frac{\partial^2H_v}{\partial\Delta_\mu\theta_1\partial\Delta_\mu\theta_2}}
      -\beta\avg{\left(\frac{\partial
            H_v}{\partial\Delta_\mu\theta_1}-\avg{\frac{\partial
              H_v}{\partial\Delta_\mu\theta_1}}\right)\left(\frac{\partial
            H_v}{\partial\Delta_\mu\theta_2}-\avg{\frac{\partial
              H_v}{\partial\Delta_\mu\theta_2}}\right)}
    \right]\label{hel_mod12}\\
    & +\frac{a_2^2}{L^3}\left[
      \avg{\frac{\partial^2H_v}{\partial\Delta_\mu\theta_2^2}}
      -\beta\avg{\left(\frac{\partial
            H_v}{\partial\Delta_\mu\theta_2}-\avg{\frac{\partial
              H_v}{\partial\Delta_\mu\theta_2}}\right)^2}
    \right]\label{hel_mod2}.
  \end{align}
  We observe that a general twist in the phases can be expressed
  through three independent quantities, the superfluid density of the
  two single components eq. \eqref{hel_mod1} and \eqref{hel_mod2},
  denoted $\Upsilon^\mu_1$ and $\Upsilon^\mu_2$ respectively, and a
  novel inter-component quantity Eq. \eqref{hel_mod12} denoted
  $\Upsilon^\mu_{12}$. We interpret $\Upsilon_{12}$ as a renormalized
  drag coefficient. A general helicity modulus may then be written in
  the compact form
  \begin{equation}
    \Upsilon^\mu_{a_1,a_2}=a_1^2\Upsilon^\mu_1 +2a_1a_2\Upsilon^\mu_{12} + a_2^2\Upsilon^\mu_2.
  \end{equation}
\end{widetext}

\section{\label{negdrag} Negative drag coefficient}
The subject of the sign of the drag coefficient, $\rho_d$ is a subtle
one and depends on the physical realization of the model. In the case of a
realization of the model on an optical lattice the sign of the drag
coefficient {  can straightforwardly be made negative \cite{Kuklov}}.

\begin{figure}[h]
  \includegraphics[width=\columnwidth]
  {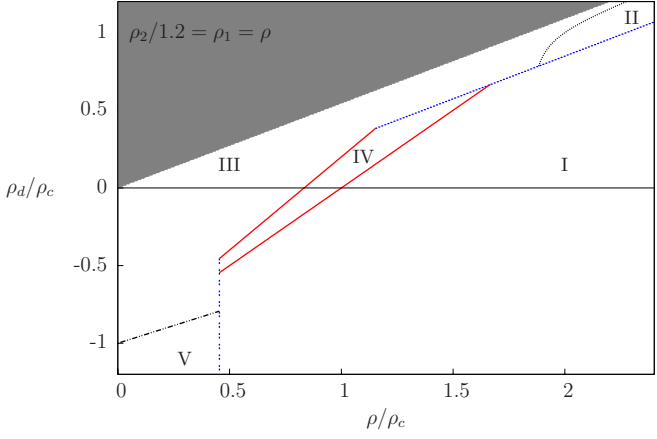}
  \caption{\label{PD4} (Color online) Same as Fig. \ref{PD2}, but now
    also including a negative drag coefficient $\rho_d$. In region I)
    the system features a broken $U(1)\times U(1)$ symmetry, in
    regions II), IV) and V) the system features a broken $U(1)$
    symmetry associated with the phase sum, one individual phase and
    the phase difference respectively. In region III) the system is in
    the fully symmetric state. (See also Fig. \ref{PD2}).}
\end{figure}

In the case of a negative drag coefficient the analysis in
Section \ref{uneqr1r2} will hold, with the role of $(1,1)$- and
$(1,-1)$-vortices interchanged. However in the separation of variables
we need to extract a phase difference to estimate the stiffness which
would remain in the system when $(1,1)$ vortices proliferate. Then,
the proper separation of variables for analyzing the model is given by
  \begin{equation}
    \begin{split}
      F=
      \int_{\boldsymbol{r}}\du\bos{r}
      \bigg\{ & 
        \left(
          \frac{\alpha}{1+\alpha} \rho - \rho_d
        \right) 
        \left[ 
          \nabla(\theta_1-\theta_2) 
        \right]^2 \\
        & + 
        \frac{\rho}{1+\alpha}  
        \left[ 
          \nabla \theta_1 +  \alpha\nabla \theta_2 
        \right]^2
      \biggr\}.        
    \end{split}
    \label{freeenergy_sIII}
  \end{equation}      

\end{document}